# Applying Multi-Core Model Checking to Hardware-Software Partitioning in Embedded Systems (extended version)

Alessandro Trindade, Hussama Ismail, and Lucas Cordeiro
Federal University of Amazonas - Manaus, Amazonas, Brazil
{alessandro.b.trindade, hussamaismail}@gmail.com, lucascordeiro@ufam.edu.br

*Abstract*—We present an alternative approach to solve the hardware (HW) and software (SW) partitioning problem, which uses Bounded Model Checking (BMC) based on Satisfiability Modulo Theories (SMT) in conjunction with a multi-core support using Open Multi-Processing. The multi-core SMT-based BMC approach allows initializing many verification instances based on processors cores numbers available to the model checker. Each instance checks for a different optimum value until the optimization problem is satisfied. The goal is to show that multi-core model-checking techniques can be effective, in particular cases, to find the optimal solution of the HW-SW partitioning problem using an SMT-based BMC approach. We compare the experimental results of our proposed approach with Integer Linear Programming and the Genetic Algorithm.

## I. INTRODUCTION

Nowadays, with the strong development of embedded systems, the design phase plays an important role. At early stages, the design is split into separated flows: hardware and software. Consequently, the partitioning decision process, which deals with the decisions upon which parts of the application have to be designed in hardware (HW) and which in software (SW), must be supported by any well-structured methodology. If not, this leads to a number of issues (design flow interruptions, redesigns, and undesired iterations) which affects the overall development process, the quality and the lifecycle of the final system. Starting at the 1990s, intensive research was performed, and several approaches proposed, as shown in [1] and [2].

In any HW and SW design of complex systems, more time is spent on verification than on construction [3]. Formal methods based on model checking offer great potential to obtain a more effective and faster verification in the design process. Programs may be viewed as mathematical objects with behavior that is, in principle, well determined. This makes it possible to specify programs using mathematical logic, which constitutes the intended (correct) behavior. Then, one can try to give a formal proof or otherwise establish that the program meets its specification [4]. Research in formal methods has led to the development of very promising verification techniques, which facilitate the early detection of errors. Model-based verification techniques use models that describe the possible system behavior in a mathematically precise and unambiguous manner. The system models are accompanied by algorithms that systematically explore all the states of the system model.

In [5] and [6] was shown that it is possible to use Bounded Model Checking (BMC) based on Satisfiability Modulo Theories (SMT) to perform HW-SW partitioning in embedded systems. The present work extends those studies since there is a substantial improvement in terms of the genetic algorithm and the SMT-based verification method, which has been extended with a multi-core architecture. Multi-core processors have been used in all segments of industry to implement high-performance computing [7]. In particular, hardware platforms, together with multi-processing platforms, have allowed verification algorithms to distribute tasks executions across multiple processors, which generate an increase in performance if compared to single-core solution. However, most verification algorithms still disregard the limitations of the CMOS technology, which limits the increase of the chip's frequency after it reaches 4 GHz.

Here, we exploit the availability of multi-core processors. In particular, a multi-core SMT-based BMC method is applied to the HW-SW partitioning and then is compared to the results with classical integer linear programming (ILP) and genetic algorithm (GA) using a multi-core tool as well. To the best of our knowledge, this is the first work to use a multi-core SMT-based verification to solve a HW-SW partitioning problem in embedded systems. We implement our ideas with the Efficient SMT-based Bounded Model Checker (ESBMC) tool [14]. As its main contribution, this paper shows that it is possible to take advantage of an SMT-based BMC tool in a multi-core architecture to solve optimization problems.

This paper is organized as follows: Section II gives a background on optimization, model checking, and multi-core support with Open Multi-Processing. Section III describes informal and formal mathematical modeling. Section IV describes briefly the binary integer programming and GA algorithms. The SMT-based BMC method is presented in Section V. Section VI presents the experimental evaluation. Section VII discusses related work. Section VIII presents the conclusion and future work.

## II. BACKGROUND

### A. Optimization

Optimization is the act of obtaining the best result (i.e., the optimal solution) under given circumstances [9]. In the design, construction, and maintenance of any engineering system, engineers have to make many technological and managerial decisions at several stages. The ultimate goal of all such decisions is either to minimize the effort required or to maximize the desired benefit. Because the effort required or the benefit desired in any practical situation can be expressed as a function of certain decision variables, optimization can be defined as the process of finding the conditions that give the maximum or minimum value of a function [9].

There is no single method available for solving all optimization problems efficiently [9]. The most known technique is linear programming, which is an method applicable for the solution of problems in which the objective function and the constraints appear as linear functions of the decision variables. A particular case of linear programming is ILP, in which the variables can assume just integer values. Eq. (1) shows a typical linear programming problem, where $A$ and $b$ are vectors or matrixes that describe the constraints.

$$min\ f^T x\ such\ that \begin{cases} A.x \leq b, \\ Aeq.x = beq, \\ x \geq 0. \end{cases} \quad (1)$$

In some cases, the time to find a solution using ILP is impractical. Even with the use of powerful computers, a problem can take hours running before an optimal solution is reached. If the optimization problem is complex, some heuristics can be used to solve the same problem faster, e.g., those used in the GA [9]. The only drawback is that the found solution may not be the global minimum or maximum.

### B. Bounded Model Checking with ESBMC

Model checking refers to algorithms for exploring the state space of a transition system to determine if it obeys a specification of its intended behavior [3],[4]. These algorithms can perform exhaustive exploration in a highly automatic way and, thus, have attracted much interest in industry. However, model-checking has been held back by the state explosion problem, in which the number of states in a system grows exponentially in the number of system components [10]. Much research has been devoted to mitigate this problem.

Among the recent techniques, there is one that combines model checking with satisfiability solving. This technique, known as bounded model checking (BMC), does a very fast exploration of the state space, and for some types of problems, it offers large performance improvements over previous approaches, as shown in [10]. In particular, BMC based on Boolean Satisfiability (SAT) has been introduced as a complementary technique to binary decision diagrams for alleviating the state explosion problem.

The basic idea of BMC is to check the negation of a given property at a given depth: given a transition system $M$, a property $\phi$, and a bound $k$, BMC unrolls the system $k$ times and translates it into a verification condition (VC) $\varphi$ such that $\varphi$ is satisfiable if and only if $\phi$ has a counterexample of depth $k$ or less [10]. To cope with increasing software complexity, SMT solvers can be used as back-ends for solving the generated VCs, as shown in [11], [12], and [13].

According to [14] and [15], SMT-based model checking can be used to verify the single- and multi-threaded software. In [16], ESBMC can also be used to model check C++ software based on SMT solvers. In [5] and [6] it was shown that it is possible to use ESBMC, as an optimization tool.

There are two directives in C/C++ that can be used to guide a model checker to solve an optimization problem: ASSUME and ASSERT. The directive ASSUME is responsible for ensuring the compliance of constraints (software costs), and the directive ASSERT controls the halt condition or code violation (minimum hardware cost). Then, with some C/C++ code, it is possible to guide ESBMC to solve optimization problems.

### C. Multi-core ESBMC with OpenMP

Nowadays, although the CPU used to perform tests usually has a modern multi-core architecture, with the ability to run several threads on different processing cores, ESBMC verification runs are still performed only in a single-core. For instance, if the processor has 8 processing cores available, only one is used for the verification and the others remain idle. There is a significant unused hardware resource during this process.

Fig.1 shows the ESBMC architecture, which consists of the C/C++ parser, GOTO Program, GOTO Symex, and SMT solver [16]. In particular, ESBMC compiles the C/C++ code into equivalent GOTO-programs (i.e., control-flow graphs) using a gcc-compliant style. The GOTO-programs can then be processed by the symbolic execution engine, called GOTO Symex, where two recursive functions compute the constraints ($C$) and properties ($P$); finally it generates two sets of equations (i.e., $C \land \neg P$) which are checked by an SMT solver.

The main factor for ESBMC to use only a single-core relies on its back-end (i.e., SMT Solver). Currently, the SMT solvers supported by ESBMC are: Z3 [24], Boolector [25], MathSAT [26], CVC4 [27], and Yices [28]. Most of them do provide neither multi-threaded support nor a parallel version to solve the generated SMT equations.

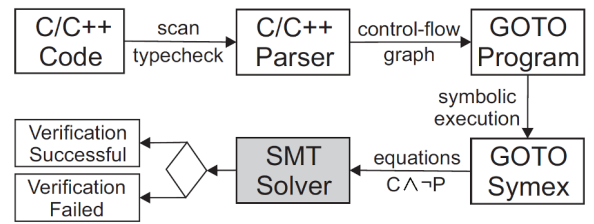

Fig. 1. ESBMC architecture

To optimize the CPU resources utilization without modifying the underlying SMT Solver, the Open Multi-Processing (OpenMP) library [23] is used in this present work as a front-end for ESBMC.

OpenMP is a standard Application Programming Interface (API) for shared memory programming, which has been very successful for structural parallelism in applications. The API provides a directive-based programming approach to write parallel versions of C/C++ programs [33]. In OpenMP, the implementation is based on the fork-join model. The main thread executes the sequential parts of the program; if a parallel region is encountered, then it forks a team of worker threads. After the parallel region finishes (i.e., the API waits until all threads terminate), then the main procedure gets back to the single-threaded execution mode [7]. Fig. 2 shows our approach called "Multi-core ESBMC".

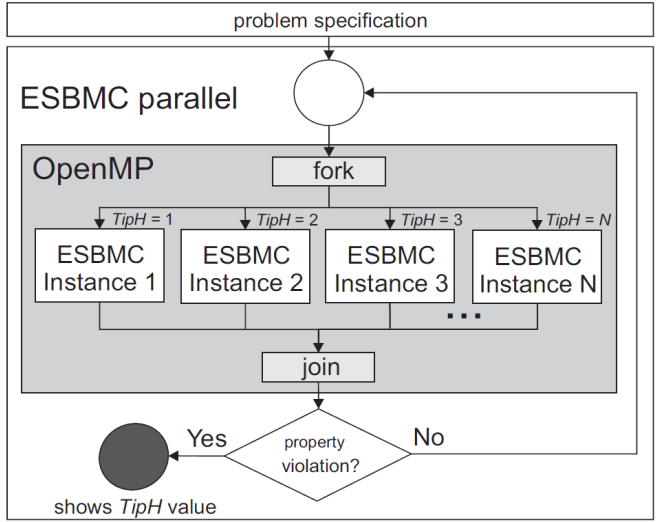

Fig. 2. Multi-core ESBMC Approach

Multi-core ESBMC obtains the problem specification represented by a C program, which is violated when the correct optimum value ($TipH$) parameter is reached; Multi-core ESBMC starts a parallel region with $N$ different instances of ESBMC, based on the number of available processing cores. All these ESBMC instances run independently of each other, as shown in Fig. 2; note that there is no shared-memory (or message-passing) mechanism among the threads. In particular, different threads are managed by the OpenMP API, which is responsible for the thread lifecycle: start, running, and dead states, using different $TipH$ values as condition. After executing $N$ instances, if there is no code violation, then multi-core ESBMC starts $N$ new instances again. During the parallel region execution, if a violation is found in any running thread, then it presents the counterexample with the violation condition and the verification time. If all threads of the batch processing are terminated, then multi-core ESBMC finishes its execution.

III. MATHEMATICAL MODELING

The mathematical modeling was taken from [1], [2].

*A. Informal Model (or Assumptions)*

The informal model can be described by five characteristics. First, there is only one software context, i.e., there is just one general-purpose processor, and there is only one hardware context. The components of the system must be mapped to either one of these two contexts. Second, the software implementation of a component is associated with a software cost, which is the running time of the component. Third, the hardware implementation of a component has a hardware cost, which can be area, heat dissipation, and energy consumption. Fourth, based on the premise that hardware is significantly faster than software, the running time of the components in hardware is considered as zero. Finally, if two components are mapped to the same context, then there is no overhead of communication between them; otherwise, there is an overhead of communication. The consequence of these assumptions is that scheduling does not need to be addressed in this work. Hardware components do not need scheduling, because the running time is assumed to be zero. Because there is only one processor, software components do not need to be scheduled as well. Therefore, the focus is only on the partitioning problem. That configuration describes a first-generation co-design, where the focus is on bipartitioning [17].

*B. Formal Model*

The inputs of the problem are: A directed simple graph $G = (V, E)$, called the task graph of the system, is necessary. The vertices $V = \{v_1, v_2, ..., v_n\}$ represent the nodes that are the components of the system that will be partitioned. The edges ($E$) represent communication between the components. Additionally, each node $v_i$ has a cost $h(v_i)$ (or $h_i$) of hardware (if implemented in hardware) and a cost $s(v_i)$ (or $s_i$) of software (if implemented in software). Finally, $c(v_i, v_j)$ represents the communication cost between $v_i$ and $v_j$ if they are implemented in different contexts (hardware or software).

Based on [1], $P$ is called a hardware-software partition if it is a bipartition of $V$: $P = (V_H, V_S)$, where $V_H \cup V_S = V$ and $V_H \cap V_S = \emptyset$. The crossing edges are $E_P = \{(v_i, v_j): v_i \in V_S, v_j \in V_H \text{ or } v_i \in V_H, v_j \in V_S\}$. The hardware cost of $P$ is given by Eq. (2), and the software cost of $P$ (i.e., software cost of the nodes and the communication cost) is given by Eq. (3):

$$H_P = \sum_{v_i \in V_H} h_i \qquad (2)$$

$$S_P = \sum_{v_i \in V_S} s_i + \sum_{(v_i, v_j) \in E_P} c(v_i, v_j) \qquad (3)$$

Three different optimization and decision problems can be defined. In this paper, the focus is on the case that $S_0$ is given, i.e., to find a $P$ HW-SW partitioning so that $S_P \leq S_0$ and $H_P$ is minimal (system with hard real-time constraints). So, based on Eq. (1) and Eq. (3) the optimization problem's restrictions can be reformulated as: $s(1 - x) + c|Ex| \leq S_0$, where $x$ is the decision variable. Concerning the complexity of this problem, reference [1] demonstrates that it is NP-Hard.

IV. PARTITIONING PROBLEM USING ILP-BASED AND GENETIC ALGORITHMS

The ILP and GA were taken from [5] and [6]. Both use slack variables in order to be possible to represent the constraints and to use commercial tools. However, GA had improvements from the parameters of related studies in order to increase the solution accuracy without producing timeout. The tuning was performed by empirical tests and resulted in changing of three parameters, which are passed to function *ga* of MATLAB [18]: the population size was set from 300 to 500, the Elite count changed from 2 (default value) to 50, and the number of Generations changed from 100*$NumberOfVariables$ (default) to 75.

## V. ANALYSIS OF THE PARTITIONING PROBLEM USING ESBMC

ESBMC pseudocode shows the algorithm with the same restrictions and conditions placed on ILP and GA. Two values must be controlled to obtain the results and to perform the optimization. One is the initial software cost, as defined in Section III.B. The other is the halting condition (code violation) that stops the algorithm.

The ESBMC algorithm starts with the declarations of hardware, software, and communication costs. $S_0$ also must be defined, as the transposed incidence matrix and the identity matrix, as typically done in MATLAB. Here, the matrices A and b are generated. At that point, the ESBMC algorithm starts to differ from the ILP and GA presented in [5] and [6].

It is possible to tell the ESBMC with which type of values the variables must be tested. Therefore, there is a declaration to populate all the decision variables $x$ with non-deterministic Boolean values. Those values that change for each test will generate a possible solution and obey the restrictions. If this is achieved, then a feasible solution is found and the ASSUME directive is responsible for ensuring the compliance of constrains (i.e., $A.x \leq b$).

A loop controls the cost of hardware hint, starting with zero and reaching the maximum value considering the case, where all nodes are partitioned to hardware. To every test performed, the hardware hint is compared to the feasible solution. This is accomplished by an ASSERT statement at the end of the algorithm, a predicate that controls the halt condition (true-false statement). If the predicate is FALSE, then the optimization is finished, i.e., the solution was found. The ASSERT statement tests the objective function, i.e., the hardware cost, and will stop if the hardware cost found is lower than or equal to the optimal solution. However, if ASSERT returns a TRUE condition, i.e., the hardware cost is higher than the optimal solution, then the model-checking algorithm restarts and a new possible solution is generated and tested until the ASSERT generates a FALSE condition. When the FALSE condition happens at verification-time, the execution code is aborted and ESBMC presents the counterexample that caused the condition to be broken. That is the point in which the solution is presented (minimum HW cost).

In the ESBMC algorithm, which is shown below, it is not necessary to add slack variables because the modulus operation is kept, which reduces the number of variables to be solved.

ESBMC Pseudocode
  Initialize variables
  Declare number of nodes and edges
  Declare hardware cost of each node as array ($h$)
  Declare software cost of each node as array ($s$)
  Declare communication cost of each edge ($c$)
  Declare the initial software cost ($S_0$)
  Declare transposed incidence matrix graph $G$ ($E$)
  Define the solutions variables ($x_i$) as Boolean
  main{
    For $TipH = 0$ to $Hmax$ do {
      Populate $x_i$ with nondeterministic/test values
      Calculate $s(1-x) + c*|Ex|$ and store at $variable$
      Requirement insured by ASSUME ($variable \leq S_0$)
      Calculate $H_P$ cost based on value tested of $x_i$
      Violation check with ASSERT ($H_P > TipH$)
    }
  }

In the multi-core ESBMC algorithm, the only difference is the fact that the value of $TipH$ and its range is not declared in the algorithm, as shown in ESBMC Pseudocode. The proposed approach is invoked for each test problem, as follows:

esbmc-parallel $<filename.c>$ $<hmin\_value>$ $<Hmax>$

Where $<filename.c>$ is the optimization problem described in ANSI-C format, $<hmin\_value>$ is the minimum (zero to HW-SW partitioning problem) and $<Hmax>$ is the maximum hardware cost for the specified problem.

Therefore, the algorithm starts $N$ different instances of ESBMC using the different optimization values, in ascending order, for $Hmax$ in order to find a violation. If all instances finish and no violation is found, then multi-core ESBMC starts new $N$ instances. When a violation is found, it reports time and hardware cost. If multi-core ESBMC tests all the possibilities for the hardware cost and has not found a violation, then it reports: "Violation not found".

## VI. EXPERIMENTAL EVALUATION

ESBMC 1.24 running on a 64-bit Ubuntu 14.04.1 LTS operating system was used. Version 2.0.1 of Boolector SMT-solver [25] (freely available) was used as well. For the ILP and GA formulations, MATLAB R2013a from MathWorks with Parallel Computing Toolbox was used [18]. MATLAB is a dynamically typed high-level language known as the state-of-the-art mathematical software [19] and is widely used by the engineering community [20]. The ESBMC multi-core algorithm was implemented in C++ [1]. A desktop with 24GB of RAM and i7 (8-cores) from Intel with clock of 3.40 GHz was used. Each time was measured 3 times in GA (average taken) and just once in ESBMC and ILP. The reason is that GA times are not so close as ESBMC and ILP. A time out condition (TO) is reached when the running time is longer than 7,200 seconds. A memory out (MO) occurs when the tool reaches 24GB of memory. TABLE I. lists the benchmarks[1].

TABLE I.     DESCRIPTION OF BENCHMARKS

| Name | Nodes | Edges | Description |
|---|---|---|---|
| CRC32 | 25 | 32 | 32-bit cyclic redundancy check [21] |
| Patricia Insert | 21 | 48 | Routine to insert values [21] |
| Dijkstra | 26 | 69 | Computer shortest paths in a graph [21] |
| Clustering | 150 | 331 | Image segmentation algorithm in a medical application |
| RC6 | 329 | 448 | RC6 cryptography graph |

---

[1] Available at: http://www.esbmc.org/benchmarks/

| Fuzzy | 261 | 422 | Clustering algorithm based on fuzzy logic |
| Mars | 417 | 600 | MARS cipher from IBM |

The vertices in the graphs correspond to high-level language instructions. Software and communication costs are time dimensional, and hardware costs represent the occupied area.

The overall performance (TABLE II. ) shows that ILP is the best solution of all techniques, even if we consider that the Fuzzy benchmark reached time out with ILP. Thus, the maximum limit to use ILP is around 329 nodes or less. GA was the only technique that could solve all benchmarks, but the error from the exact solution varied from -37.6% to 29%.

Multi-core ESBMC had a better performance than that of pure ESBMC. The relative speedup obtained ranged from 1.9 to 60.3, which shows a reasonable improvement. Until the number of 150 nodes is reached, the ESBMC technique, mainly Multi-core ESBMC, has shown itself to be a good choice to solve HW-SW partitioning. This is because the exact solution was found and the execution time was mostly closer to ILP (from the same performance to 4.7 times faster). If the complexity of test vectors increases, then pure ESBMC algorithm has the drawback of creating an even more complex problem, because it increases the states created, which controls the hardware cost hint.

TABLE II. RESULTS OF THE BENCHMARKS

|  |  | CRC32 | Patricia | Dijkstra | Clustering | RC6 | Fuzzy | Mars |
|---|---|---|---|---|---|---|---|---|
|  | Nodes | 25 | 21 | 26 | 150 | 329 | 261 | 417 |
|  | Edges | 32 | 48 | 69 | 331 | 448 | 422 | 600 |
|  | $S_0$ | 20 | 10 | 20 | 50 | 600 | 4,578 | 300 |
| Exact Solution | $H_P$ | 15 | 47 | 31 | 241 | 692 | 13,820 | 876 |
|  | $S_P$ | 19 | 4 | 19 | 46 | 533 | 4,231 | 297 |
| ILP | Time (s) | 2 | 1 | 2 | 649 | 1,806 | TO | 5,429 |
|  | $H_P$ | 15 | 47 | 31 | 241 | 692 | - | 876 |
| GA | Time (s) | 7 | 7 | 9 | 340 | 2,050 | 1,372 | 5,000 |
|  | Error % | 13.3 | 0.0 | 29.0 | 1.7 | -6.5 | -37.6 | -27.5 |
| ESBMC | Time s | 31 | 362 | 292 | 3,010 | TO | MO | MO |
|  | $H_P$ | 15 | 47 | 31 | 241 | - | - | - |
| Multi-core ESBMC | Time (s) | 2 | 6 | 7 | 1,615 | TO | TO | TO |
|  | $H_P$ | 15 | 47 | 31 | 241 | - | - | - |
| ESBMC Relative Speedup |  | 15.4 | 60.3 | 41.7 | 1.9 | - | - | - |

Legend: TO = Time out and MO = memory out

With the RC6 benchmark (329 nodes), ESBMC was unable to present a solution without exceeding the time limit of 7200 seconds. Pure ESBMC had even a worse performance with Fuzzy and Mars, because the execution presented memory out. This is a clear indication that the prune method adopted by the ILP's search tree solver is still more efficient than that adopted by ESBMC solver.

## VII. RELATED WORK

Since the second half of the first decade of the 2000s, three main paths have been tracked to improve or to present alternative solutions to the optimization of HW-SW partitioning, i.e., to find the exact solution [2], to use heuristics to speed up performance time [1], and hybrid ones [22].

In the first group, the exact solution to the HW-SW partitioning problem is found. The use of SMT-based verification presented in this paper can be grouped into this category, because the exact solution is found with the given algorithm. The difference is based only in terms of the technique chosen to solve the problem.

Another path followed in past initiatives and which has had more studies is the creation of heuristics to speed up the running time of the solution. The difference between this kind of solutions and SMT-based verification is based on two facts: ESBMC is guaranteed to find the exact solution, but the heuristics are faster when the complexity is greater.

Finally, there are approaches that mixes heuristics with exact solution tools. The idea is to use a heuristic to speed up some phase of an exact solution tool. It worth mentioning that the final solution is not necessarily an optimal global solution. Only the SMT-based verification is guaranteed to find the exact solution, but hybrid algorithms are faster when complexity rises.

In terms of SMT-based verification, most work is restricted to present the model, its modification to programming languages (e.g., C/C++ and Java), and the application to multi-thread algorithms or to embedded systems to check for program correctness. In [16] it presents a bounded model checker for C++ programs, which is an evolution of dealing with C programs and [14] uses the ESBMC model checker for embedded ANSI-C software. In [5] and [6] it was proven that it is possible to use ESBMC to solve HW-SW partitioning, but in a single core way. There are related studies focused on decreasing the verification time of model checkers by applying Swarm Verification [29], and modifications of internal search engines to support parallelism [30], but there is still the need for initiatives related to parallel SMT solvers [31]. Recently, the SMT solver Z3 has been extended to pose and solve optimization problems modulo theories [32].

## VIII. CONCLUSIONS

Concerning the comparative tests, with the four techniques presented in this paper to solve HW-SW partitioning, it was evident that none of them is indicated to partition problems with more than 400 nodes. The computing time to solve the optimization problem reached some hours of execution on a standard desktop computer. If we consider less than 400 nodes, then it is possible to use ILP as the best solution provider. If the problem to be solved has 150 nodes or less, then ESBMC represents a feasible alternative. GA had an intermediate result in terms of performance, but the error presented from exact solution made it not acceptable to that kind of application. This error may be reduced by changing some parameters.

If considering off-the-shelf tools, as MATLAB to ILP and GA, the coding is simpler. However, ESBMC has a BSD-style license and can be downloaded and used for free. Concerning the two versions of ESBMC, it is possible to conclude that Multi-core ESBMC had better performance results than pure ESBMC. Thus, considering that nowadays the processors have more and more cores, when modeling the problem, it is possible to consider multi-core ESBMC as an alternative to solve the partitioning problem. Future work can be done to decrease the processing time of ESBMC (solver included).

Finally, there is an issue about 150 nodes problem, since it seems to be the limit of ESBMC. It really depends on the modeling granularity of the problem. Some researchers propose fine-grained models, in which each instruction can be mapped to either HW or SW. This may lead to thousands of nodes or even more. Others defend coarse-grained models, where decisions are made for bigger components, thus even complex systems may consist of just some dozens of nodes to partition. In principle, a fine-grained approach may allow to obtain better partitions, but at the cost of an exponential increase of the size of the search space. In future work, we will exploit other search strategies in the multi-core ESBMC approach and address specifically more complex types of architectures in the HW-SW partitioning problem, including more than one CPU.